\documentclass[runningheads]{llncs}
\usepackage[T1]{fontenc}
\usepackage{graphicx}
\usepackage{xcolor}

\usepackage{amsmath}
\usepackage{amssymb}
\usepackage{framed}

\begin{document}

\title{Topological Analysis of Seizure-Induced Changes in Brain Hierarchy Through Effective Connectivity}

\titlerunning{Topology of Effective Brain Connectivity}


\author{%
  Anass B.~El-Yaagoubi\inst{1}\orcidID{0000-0002-8886-0587} \and
  Moo K.~Chung\inst{2}\orcidID{0000-0003-2852-9670} \and
  Hernando ~Ombao\inst{1}\orcidID{0000-0001-7020-8091}
}
\authorrunning{A. El-Yaagoubi et al.}
%
\institute{
  King Abdullah University of Science and Technology (KAUST), Statistics Program, Thuwal, 23955, Saudi Arabia\\
  \email{anass.bourakna@kaust.edu.sa; hernando.ombao@kaust.edu.sa} \and
  University of Wisconsin-Madison, Department of Statistics, Medical Science Center 4725, 1300 University Ave, Madison, WI 53706, USA\\
  \email{mkchung@wisc.edu}
}

\maketitle              
\begin{abstract}
    Traditional Topological Data Analysis (TDA) methods, such as Persistent Homology (PH), rely on distance measures (e.g., cross-correlation, partial correlation, coherence, and partial coherence) that are symmetric by definition. While useful for studying topological patterns in functional brain connectivity, the main limitation of these methods is their inability to capture the directional dynamics - which is crucial for understanding effective brain connectivity. We propose the Causality-Based Topological Ranking (CBTR) method, which integrates Causal Inference (CI) to assess effective brain connectivity with Hodge Decomposition (HD) to rank brain regions based on their mutual influence. Our simulations confirm that the CBTR method accurately and consistently identifies hierarchical structures in multivariate time series data. Moreover, this method effectively identifies brain regions showing the most significant interaction changes with other regions during seizures using electroencephalogram (EEG) data. These results provide novel insights into the brain's hierarchical organization and illuminate the impact of seizures on its dynamics.

\end{abstract}

\section{Introduction}
\label{sec:introduction}

Over the past two decades, Topological Data Analysis (TDA) has become a valuable tool in various fields, providing new insights into complex data structures~\cite{PI,PL_FIRST,BARCODES_FIRST,EDELSBRUNNER_HARER,TDA_EDELSBRUNNER}. In biology, TDA has helped clarify genetic and evolutionary processes~\cite{TDA_RABADAN}, while in materials science, it has improved predictions of properties in crystalline compounds~\cite{TDA_MATERIAL_SCIENCE}. In finance, TDA has offered a novel perspective on the dynamics of financial markets during crises~\cite{TDA_FINANCE}. In neuroscience, the use of Persistent Homology (PH), a key technique within TDA, has revealed new aspects of the topological organization of brain dependence networks. By analyzing multivariate brain signals, researchers have uncovered complex patterns and structures that deepen our understanding of brain connectivity and functionality~\cite{TDA_BRAIN_ARTERY,SPECTRAL_TDA_ANASS,TDA_BRAIN}.

However, traditional uses of PH often involve Vietoris-Rips filtrations from {\it undirected} brain dependence networks. A significant limitation of this approach is its inability to capture the directionality of connectivity between nodes in a brain network~\cite{ORIENTED_TDA_ANASS}. It is well-established that node dependencies within a brain network exhibit asymmetry, where the influence exerted by node $A$ on node $B$ can differ significantly from the influence exerted by node $B$ on node $A$. This asymmetry is crucial for understanding effective brain connectivity, which involves directional information flow, from one region into another, that often varies markedly in its reverse path, thus delineating specific neural pathways~\cite{GC_DCM,KARL_FRISTON,GC_DIRECTED_DEPENDENCE,GC_NEUROSCIENCE}.

In the study of epilepsy, understanding changes in brain connectivity during seizures is crucial. Research indicates that epileptic seizures is associated with significant alterations in interactions between brain regions in a network. These changes vary depending on whether the seizure is generalized or originates from a specific focal region and spreads across the brain's hemispheres~\cite{EPILEPSY_INTRO,EFFECTIVE_CONNECTIVITY_EPILEPSY}. Traditional TDA methods, which primarily assess functional connectivity, often fall short in accurately characterizing these dynamic changes in effective brain connectivity.

The concept of causal inference refers to learning a causal model from observational data. In cognitive neuroscience, this process is fundamental for understanding how different brain regions interact, and how these interactions may be altered due to brain disorders ~\cite{CAUSALITY_COGNITIVE_NEUROSCIENCE}. Experimental interventions (e.g., optogenetics) are considered the gold standard in causal discovery~\cite{CAUSALITY_PEARL,CAUSALITY_COGNITIVE_NEUROSCIENCE}, as they provide a way to experimentally manipulate activity of a neuron (or subpopulation of neurons) and investigate how these manipulations change the interaction of a neuron with another neuron (or another subpopulation of neurons). Through these experimental manipulations, one can gain definitive evidence of causal physiological relationships. 
While such interventions and manipulations are conducted in non-human animal studies, these are ethically and practically challenging when studying brain connectivity in humans. Consequently, we will only focus on identifying causal associations and quantifying their strength from observational data. 

To gain better understanding of the dynamic changes in brain connectivity during epileptic seizures, we have developed the Causality-Based Topological Ranking (CBTR) approach. This novel method integrates Causal Inference (CI) and Hodge Decomposition (HD), allowing for the identification of causal pathways and the \underline{hierarchical ranking} of brain regions based on their mutual influence over time. CBTR assesses dependencies conditioned on parent nodes, thereby avoiding confounding effects among observed variables. Our analysis of EEG data from subjects who experienced seizures helped us identify the brain regions most significantly impacted by seizures for each subject.

\section{Methods}
\label{sec:method}

The goal is to develop a method for ranking brain regions (derived from the network of electroencehphalogram data) according to effective connectivity using a two-step process. First, causal inference techniques are employed to map the hierarchical structure of brain connectivity, with a focus on the effects of seizures. Next, Hodge Decomposition is applied to net flow of information between brain regions, quantifying their mutual influence.

\subsection{Causal Inference of the Brain's Hierarchical Structure}
\label{subsec:hierarchical_structure}

\begin{figure}
    \centering
    \includegraphics[width=.9\linewidth]{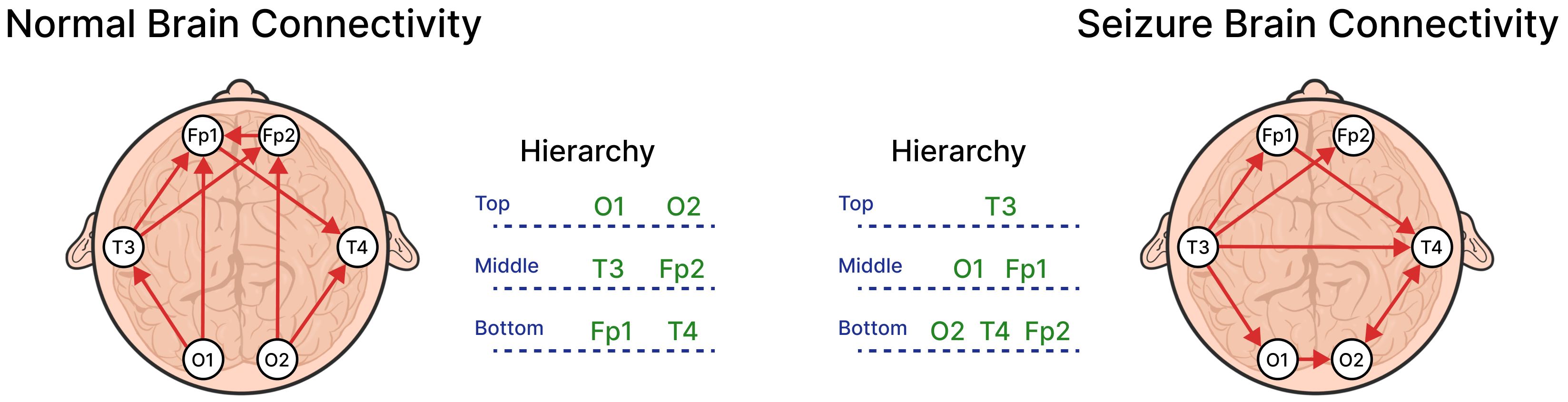}
    \caption{Potential hierarchy of the brain in normal (Left) and seizure (Right) conditions. In the normal state, the hierarchy is O1 (left occipital), O2 (right occipital) followed by T3 (left temporal), Fp2 (right frontal pole), and then T4 (right temporal), Fp1 (left frontal pole). During a seizure, the order changes to T3, O1, Fp1 followed by T4, O2, Fp2.}
    \label{fig:hierarchical_brain_connectivity_illustration}
\end{figure}

Understanding the dynamics of the hierarchical structure of the brain's effective connectivity is crucial for comprehending the impact of seizures. In an EEG network, a channel is considered to be at the top of the hierarchy if it tends to exert greater influence on other nodes, whereas a brain channel is at the bottom if it tends to be influenced more by other nodes. This hierarchical structure is determined by analyzing the causal relationships and time-delayed dependencies among the channels. This is illustrated in Figure~\ref{fig:hierarchical_brain_connectivity_illustration}. Assessing changes in brain hierarchy allows us to quantify the effects of seizures on each specific EEG channel, providing deeper insights into how seizures alter brain connectivity.

To estimate effective brain connectivity, we employ the PCMCI (Peter and Clark Momentary Conditional Independence) method, a robust causal inference technique well-suited for high-dimensional time series data \cite{PCMCI_B,PCMCI}. Following the PCMCI method, we conducted a two-stage process. First, we used the PC algorithm to identify potential causal parents for each variable, reducing dimensionality and producing a preliminary causal graph. Then, we applied the Momentary Conditional Independence (MCI) test to rigorously assess these dependencies' significance, considering both contemporaneous and time-lagged relationships while controlling for false positives. This combination of dimension reduction and rigorous testing allowed PCMCI to effectively estimate the connectivity between observed time series components, providing a detailed map of directional dependencies in the brain.

The outputs of the PCMCI method include a causal graph that highlights the most significant causal links between brain regions at various lags, with each link annotated by its corresponding statistical significance and strength. This causal graph forms the foundation for our subsequent analyses. We estimate effective connectivity as a weighted directed network, where the weight of each link ($\mathbf{W}_{p, q}$) between regions $p$ and $q$ is derived from the significance level (p-value) associated with the causal interaction from $q$ to $p$, as defined in Equations~\ref{eq:causal_dependence}, \ref{eq:mean_pvalue}.
\begin{align}
    \mathbf{W}_{p, q} &= 1 - \overline{p}_{p, q}, \label{eq:causal_dependence} \\
    \overline{p}_{p, q} &= \frac{1}{K}\sum_{k=0}^K p_{p, q}(k), \label{eq:mean_pvalue}
\end{align}
where $p_{p, q}(k)$ is the p-value corresponding to the MCI test from variable $q$ to $p$ at lag $k$. The estimated effective brain connectivity network is decomposed into symmetric ($\mathbf{W}_{s}$) and anti-symmetric ($\mathbf{W}_{a}$) components. 
\begin{align}
    \mathbf{W} &= \mathbf{W}_{s} + \mathbf{W}_{a} \label{eq:network_decomposition}
\end{align}
where $\mathbf{W}_{s} = \frac{1}{2} (\mathbf{W} + \mathbf{W}')$ represents mutual causal influence and $\mathbf{W}_{a} = \frac{1}{2} (\mathbf{W} - \mathbf{W}')$ represents net causal influence. Since brain dependencies can be asymmetric, this decomposition is relevant as it separates the connectivity network into meaningful symmetric and anti-symmetric components. The symmetric component, $\mathbf{W}_{s}$, captures bidirectional interactions, reflecting mutual influences critical for cooperative neural processes. The anti-symmetric component, $\mathbf{W}_{a}$, highlights the predominant direction of causal influence, revealing the hierarchical structure and directional information flow within the brain network. This approach provides a nuanced understanding of neural activities and identifies key regions driving effective connectivity. This decomposition is attractive to neuroscientists - it is an essential tool to  deepen our understanding of the hierarchical structure of the brain. The next step involves building the ranking of brain regions using Hodge Decomposition, which will allow us to quantify and rank the hierarchical influence of each region within the network.

\subsection{Our Proposed Method for Ranking Brain Regions}
\label{sec:ranking_brain_regions}

\begin{figure}
    \centering
    \includegraphics[width=.9\linewidth]{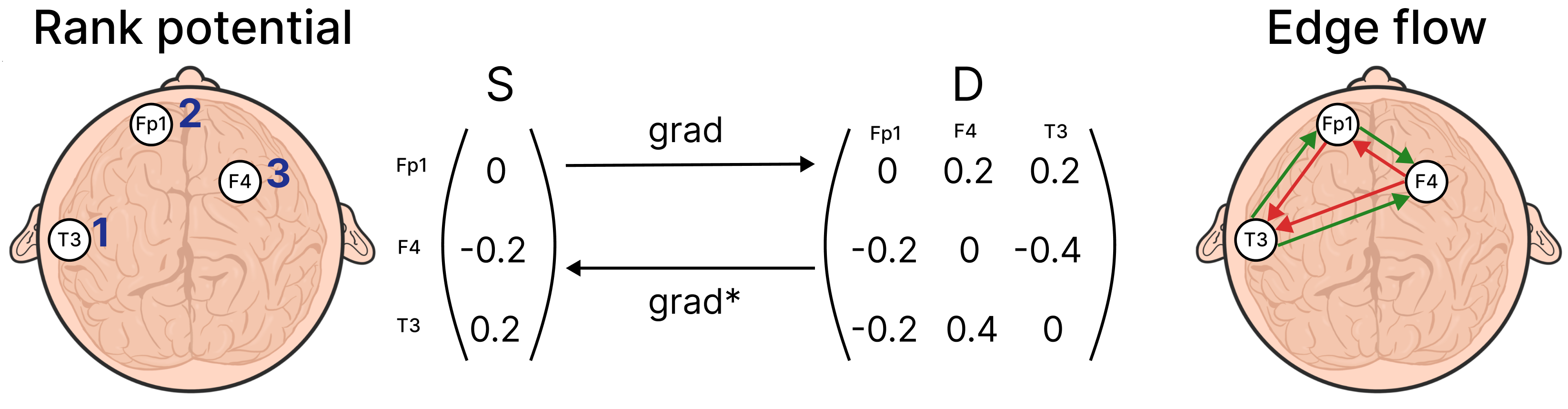}
    \caption{Node ranking in a graph involves a gradient operator that maps ranking scores on the left to gradient flow on the right (dependence structures). The adjoint, gradient star, reverses this mapping from right to left.}
    \label{fig:ranking_toy_example}
\end{figure}

Once effective connectivity is estimated and the net influence $\mathbf{W}_{a}$ is derived, our proposed method will rank brain regions by assigning a score to each node or brain channel based on their influence within the network. Our method is based on the HodgeRank approach, as described in \cite{HODGE_RANK}, which has previously been utilized to rank entities such as movies and football teams. A higher ranking score indicates a region that influences other regions more than it is influenced, while a lower score suggests a region that is more influenced by others. This concept is illustrated in Figure~\ref{fig:ranking_toy_example}.

To achieve the most meaningful ranking (i.e., one that minimizes violations of mutual net causal influence), we apply Hodge Decomposition to the anti-symmetric component of the connectivity network. Since $\mathbf{W}_{p,q} = -\mathbf{W}_{q,p}$, $\mathbf{W}_{a}$ is referred to as an alternating function. In the following, we first provide definitions for the gradient (grad) and curl operators, and then introduce the Hodge Decomposition Theorem.

\begin{definition}[Gradient Operator]
The 'grad' operator measures the flow of influence between nodes. For any edge $e = (p, q) \in E$, it is defined as:
\begin{align}
    \text{grad}(s) = s_q - s_p, \label{eq:grad_operator}
\end{align}
where $s_p$ and $s_q$ represent the scores or potentials of nodes $p$ and $q$. This operator reflects the change in influence from node $p$ to node $q$.
\end{definition}

\begin{definition}[Curl Operator]
The 'curl' operator measures the rotation strength of mutual influence among nodes. For any face $t = (p, q, r) \in T$, it is defined as:
\vspace{-2mm}
\begin{align}
    \text{curl}(\phi) = \phi_{p, q} + \phi_{q, r} + \phi_{r, p}, \label{eq:curl_operator}
\end{align}
where $\phi_{p, q}$, $\phi_{q, r}$, and $\phi_{r, p}$ are the edge flows between nodes $p, q$, and $r$. This operator evaluates the total rotation around a triangular face in the graph.
\end{definition}

Given the definitions above, we can observe that the image of the gradient operator is included in the kernel of the curl operator. This relationship leads to the Helmholtz-Hodge Decomposition theorem~\cite{HELMOLTZ_HODGE_DECOMPOSITION}.

\begin{theorem}[Helmholtz-Hodge Decomposition] \label{thm:helmholtz_hodge_decomposition}
    The space of anti-symmetric matrices, $\mathbb{R}^{E}$, can be decomposed into three mutually orthogonal subspaces:
    \begin{align*}
        \mathbb{R}^{E} = \text{Im}(\text{grad}) \oplus \text{Im}(\text{curl}^*) \oplus \text{Ker}(\text{grad}^*) \cap \text{Ker}(\text{curl}).
    \end{align*}
\end{theorem}

The anti-symmetric component $\mathbf{W}_{a}$ can exhibit cyclic influence in the network, which indicates that no perfect ranking can be achieved without violating at least one mutual influence relationship within each cycle. According to the Helmholtz-Hodge Decomposition theorem, any edge flow $\mathbf{W}_{a}$ can be uniquely decomposed as:
\vspace{-2mm}
\begin{align}
    \mathbf{W}_{a} = -\text{grad}(s) + \text{curl}^*(\phi) + \mathbf{h}, \label{eq:hodge_decomposition}
\end{align}
where $\text{grad}(s)$ is the gradient component capturing hierarchical relationships, $\text{curl}^*(\phi)$ is the curl component indicating local cyclic influences, and $\mathbf{h}$ is the harmonic component indicating non-local or global cyclic influences.

To estimate these components, we use the least squares method for the gradient and curl components, and obtain the harmonic component as the residual.
\begin{align}
    \hat{s} &= \arg \min_{s} \| \mathbf{W}_{a} + \text{grad}(s) \|_{F}^2, \\
    \hat{\phi} &= \arg \min_{\phi} \| \mathbf{W}_{a} - \text{curl}^*(\phi) \|_{F}^2, \\
    \hat{h} &= \mathbf{W}_{a} - \text{grad}(s) - \text{curl}^*(\phi).
\end{align}
The estimated score vector $\widehat{s}=[s_1, s_2, \ldots, s_N]'$ (where $N$ is the number of nodes or time series components) derived from $\mathbf{W}_{a}$ is used to establish the topological ranking of the EEG channels (or brain regions). By minimizing the difference between $-\text{grad}(\widehat{s})$ and $\mathbf{W}_{a}$, this ranking accurately reflects net information flow patterns while avoiding inconsistencies, as $\text{curl}(\text{grad}(\widehat{s})) = 0$. If $s_p > s_q$, the $p$-th time series tends to lead other time series more than the $q$-th time series, indicating that $p$ is higher in the hierarchy. Thus, $\widehat{s}$ captures the brain hierarchy, as illustrated in Figure~\ref{fig:ranking_toy_example}. This is crucial for understanding brain function and the impact of dysfunctions such as seizures.

\begin{figure}
    \centering
    \includegraphics[width=.9\linewidth]{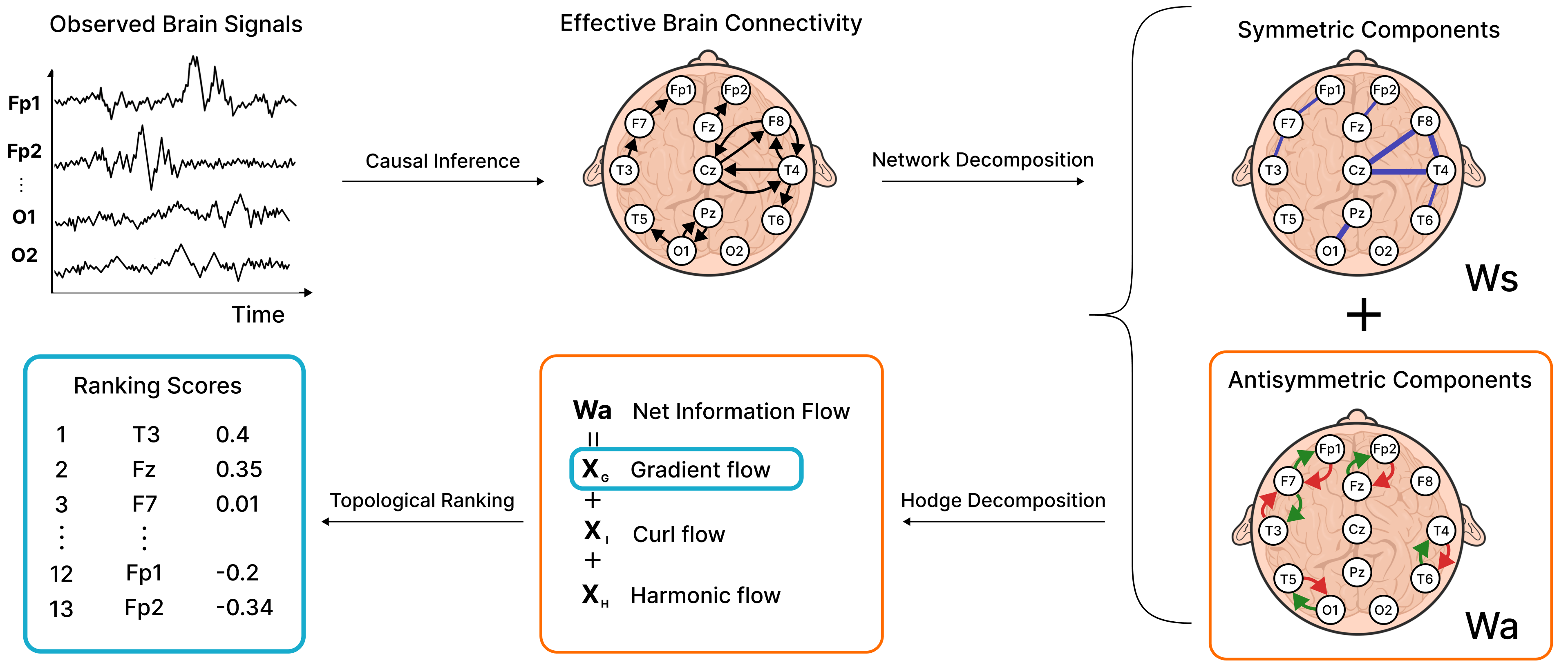}
    \caption{Hodge decomposition pipeline for topological ranking of brain regions.}
    \label{fig:hodge_pipeline}
\end{figure}

Our methodology involves a series of steps designed to rank brain regions based on their mutual influence. Initially, we employ the PCMCI method to estimate effective brain connectivity, utilizing causal inference techniques. Following this, the connectivity network is decomposed into symmetric and anti-symmetric components. We then apply Hodge Decomposition to the anti-symmetric component, extracting the gradient, which is essential for establishing the ranking. This allows us to derive the topological ranking of brain regions based on the gradient component. The entire process is visually represented in Figure~\ref{fig:hodge_pipeline}.

\vspace{-9mm}
\section{Experiments}
\label{sec:experiments}
\vspace{-2mm}

\begin{figure}
    \centering
    \includegraphics[width=.8\linewidth]{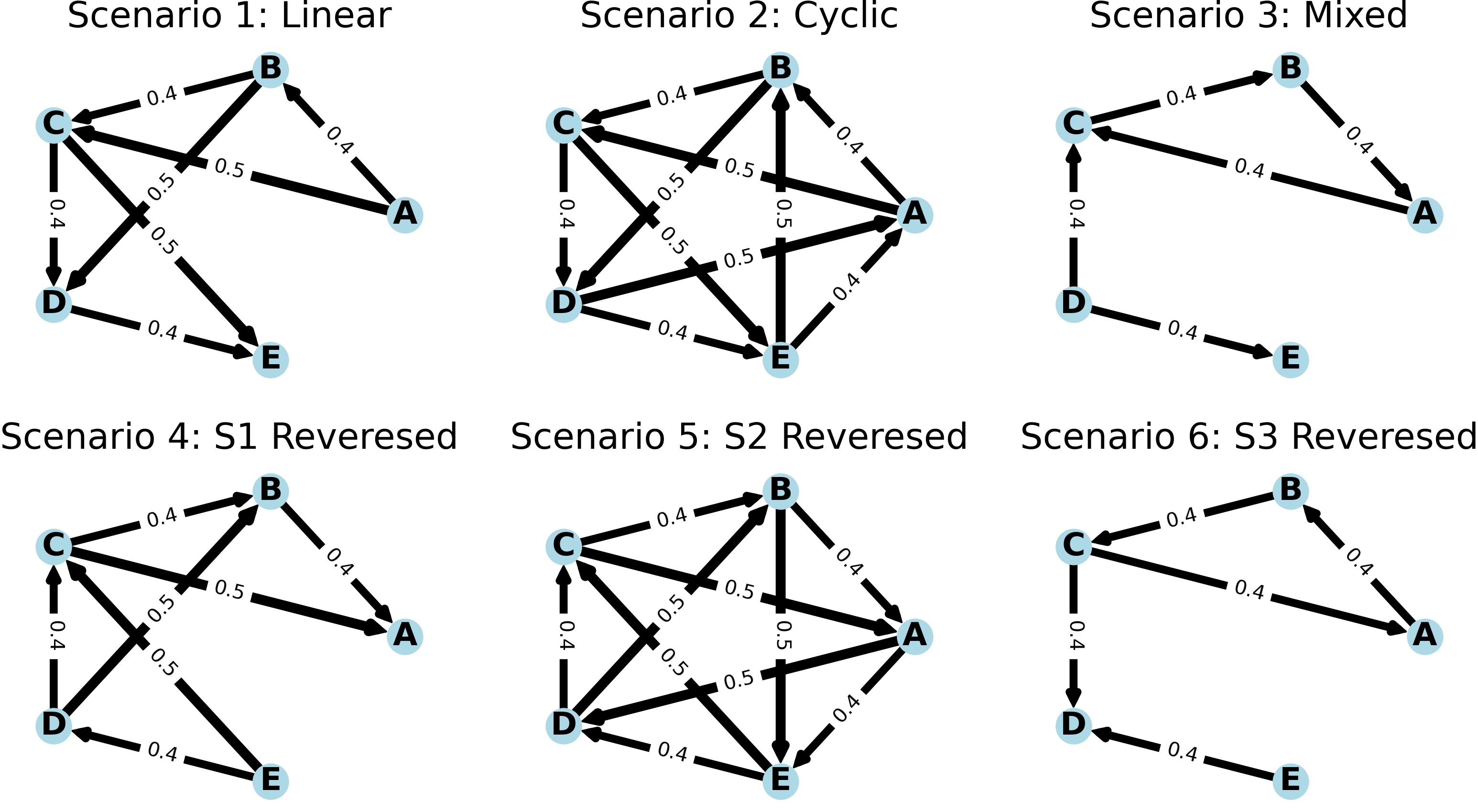}
    \caption{Visual representation of the designed dependency networks for six simulation scenarios. Each arrow indicates a dependency direction at lag one.}
    \label{fig:simulation_scenarios}
\end{figure}

To evaluate our methodology, we propose six distinct simulation scenarios, each representing different hierarchical dependency structures. These scenarios range from linear hierarchies to complex cyclic dependencies, testing the robustness and versatility of our ranking mechanism across various topologies, as illustrated in Figure~\ref{fig:simulation_scenarios}.
\begin{itemize}
    \item \textbf{Scenarios 1 and 4:} Exhibit a linear hierarchical dependence structure, allowing for clear node ranking (scenario 4 reverses scenario 1).
    \item \textbf{Scenarios 2 and 5:} Feature cyclic dependence structures, challenging meaningful ranking due to non-transitivity (scenario 5 reverses scenario 2).
    \item \textbf{Scenarios 3 and 6:} Present partial rankings with localized cycles, suggesting intermediate dependence structures (scenario 6 reverses scenario 3).
\end{itemize}
\vspace{-5mm}
\vspace{-5mm}
We generate multivariate time series data using the mixture model in Equation \ref{eq:simulations_definition}, with $P=5$ dimensions and $T=30,000$ observations, divided into 500-observation epochs to analyze temporal stability and variability in ranking. The first 5,000 observations correspond to scenario one, the next 5,000 to scenario two, and so on.
\begin{align}
    Y(t) = \Phi(t) Z(t) + E(t), \quad 
    \Phi(t) = 
    \begin{bmatrix}
        0 & 0.4 & 0.5 & 0 & 0 \\
        0 & 0 & 0.4 & 0.5 & 0 \\
        0 & 0 & 0 & 0.4 & 0.5 \\
        0 & 0 & 0 & 0 & 0.4 \\
        0 & 0 & 0 & 0 & 0 \\
    \end{bmatrix}, \label{eq:simulations_definition}
\end{align}
where $Y(t)$ is the observed multivariate time series, $Z(t)$ is a multivariate standard Gaussian process, and $E(t)$ is random noise, also modeled as a standard Gaussian process. The mixing matrix $\Phi(t)$ reflects each scenario's dependence structure.

\begin{figure}
\centering
    \includegraphics[width=.7\linewidth]{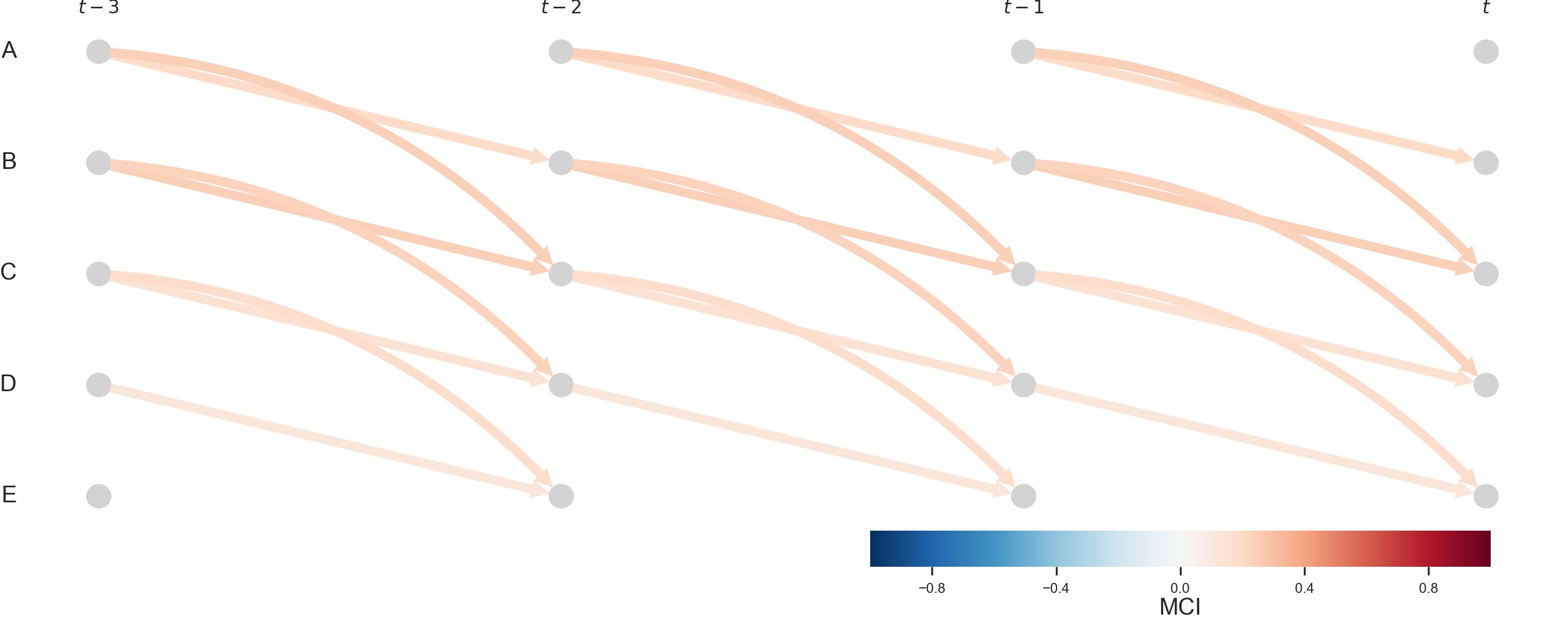}
    \caption{Visual representation of the causal interactions within the network for scenario one as identified by PCMCI, showcasing the algorithm's capacity to accurately capture both the direction and magnitude of causal dependencies across time lags.}
    \label{fig:simulations_lag_dependence}
\end{figure}

Our method follows an epoch-based approach to dynamically estimate causal relationships within the multivariate time series data in order to capture the directional dependencies and hierarchical variations. Figure \ref{fig:simulations_lag_dependence} showcases the causal connections identified for one epoch from scenario one, highlighting both the causal pathways and the strength of these interactions.

\begin{figure}
    \centering
    \includegraphics[width=.7\linewidth]{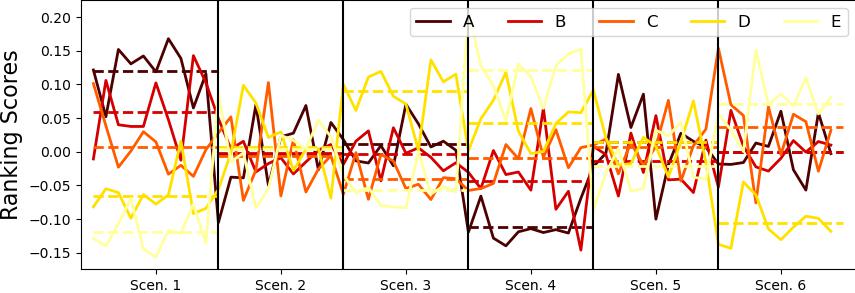}
    \caption{Aggregate causality-based ranking scores, delineating the hierarchical structure within each scenario. The distinct patterns observed across scenarios validate the robustness of our ranking methodology.}
    \label{fig:simulations_ranking_scores}
\end{figure}

By applying our proposed ranking procedure, we are able to identify and highlight nodes within the network that exert significant influence over others, providing a clear understanding of the hierarchical patterns and their dynamics, as can be seen in Figure \ref{fig:simulations_ranking_scores}. Our methodology effectively ranks the simulated time series, consistently placing node A at the top and node E at the bottom, as indicated by their respective curves, in scenario one, with the inverse pattern observed in scenario four. A node is ranked above another if its curve is consistently above the other's. Scenarios two and five exhibit fluctuating rankings due to inherent cyclic dependencies, while scenarios three and six show partial rankings with notable positions for node D, aligning with the designed simulation scenarios.
These results validate our approach for ranking multivariate time series data based on net influence, as it robustly identifies the hierarchical structures in scenarios one and three. In contrast, scenarios two and four, characterized by cyclic dependencies, exhibit inconsistent rankings. Through these simulations, we confirm that our CBTR method effectively uncovers the hierarchical structure within dependence networks, distinguishing various dependency patterns.

\vspace{-2mm}
\section{Results}
\label{sec:results}
\vspace{-2mm}

In this section, we focus on applying our methodology to a dataset of neonatal EEG recordings \cite{NEONATAL_EEG}. Neonatal seizures are a critical concern within the neonatal intensive care unit (NICU), raising numerous questions about their temporal and spatial characteristics. The dataset comprises 19-channel EEG recordings from 79 full-term infants at the Helsinki University Hospital NICU, with seizure annotations provided by three expert clinicians for each recording, spanning a median duration of 74 minutes. We concentrate our study on a few subjects, specifically analyzing epochs consistently classified as either seizure or seizure-free by all three experts. Our initial goal is to explore how seizures influence effective brain connectivity in neonates, serving as a foundational step for future detailed research on the nature of seizures and their developmental effects.

\begin{figure}
    \centering
    \includegraphics[width=.7\linewidth]{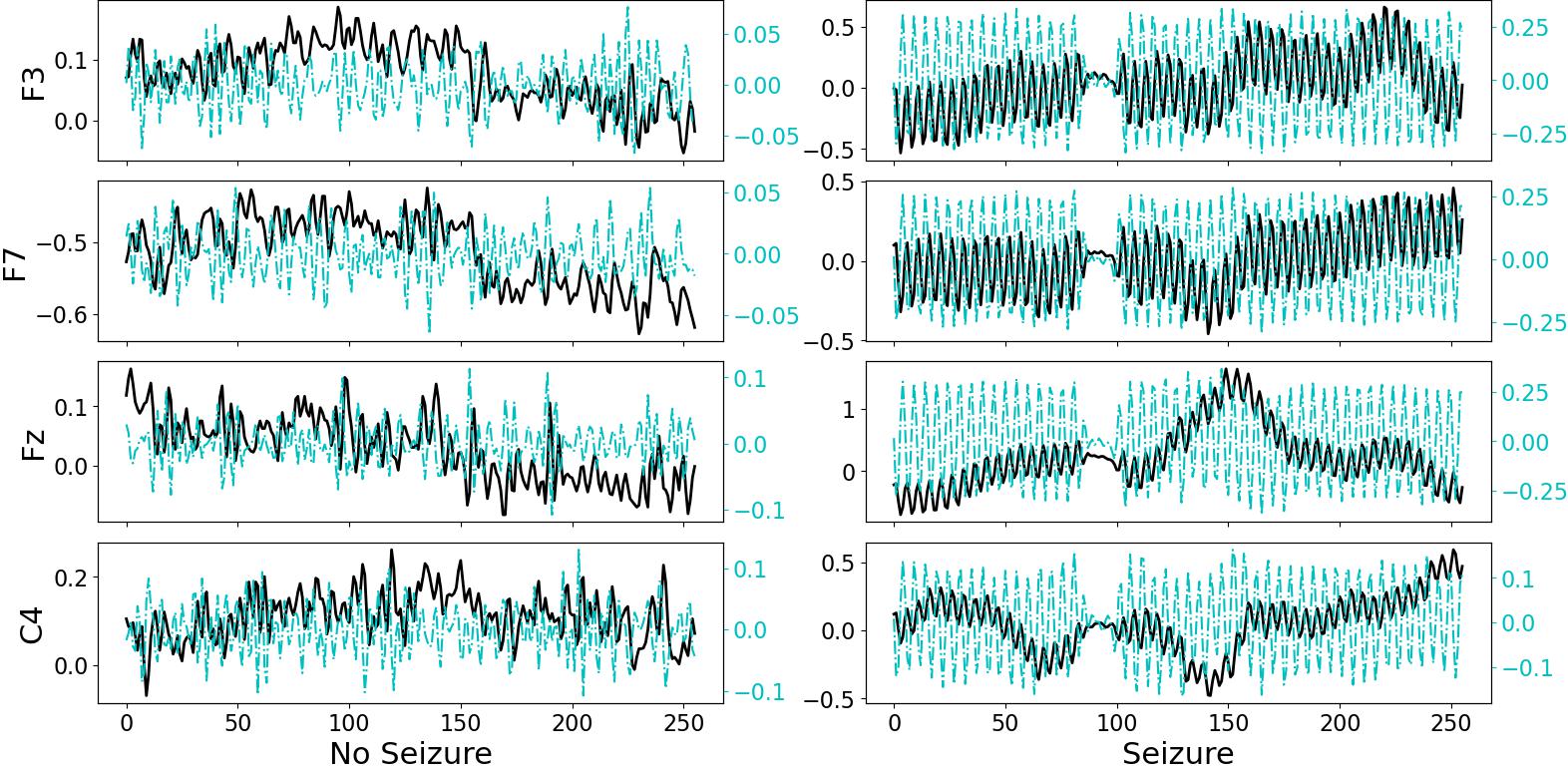}
    \caption{Sample of EEG signals from channels F3, F7, Fz, and C4, showcasing epochs (1 second each) with and without seizure. Original EEG recordings are depicted in black, while the incremental changes between successive EEG readings are highlighted in cyan.}
    \label{fig:neonatal_eeg}
\end{figure}

In many instances, neonatal EEG recordings reveal distinct seizure patterns, as shown in Figure \ref{fig:neonatal_eeg}. However, deciphering the interconnections among the 19 channels, particularly the influence of seizures, presents a complex challenge. Our approach quantifies the impact of seizures on the neonatal brain's effective connectivity at the channel level, aiming to uncover how seizures modify brain connectivity and hierarchical organization in each subject.
The EEG recordings typically exhibit smooth variations at the millisecond scale, leading to high autocorrelation in lagged observations, which can increase spurious causal relationships. To mitigate this, we analyze effective brain connectivity through successive differences or increments in EEG recordings, reducing the influence of lagged dependencies, as illustrated in Figure \ref{fig:neonatal_eeg}.

\begin{figure}
    \centering
    \includegraphics[width=.9\linewidth]{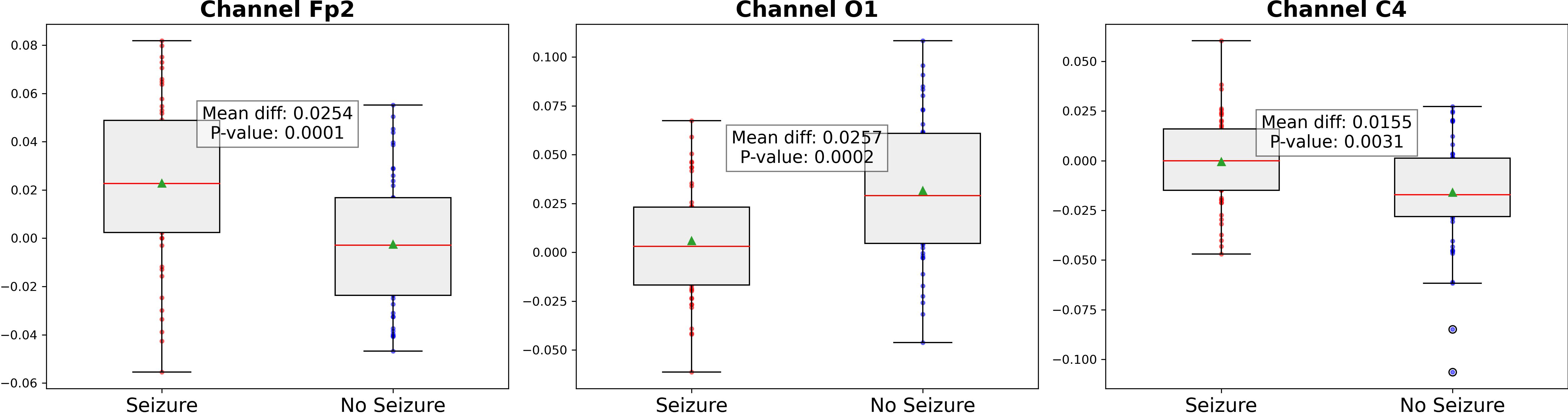}
    \caption{Channels with most significant shifts in the ranking scores between seizure free epochs and seizure epochs. For subject 1.}
    \label{fig:eeg_ranking_dynamics_subject_1}
\end{figure}

\begin{figure}
    \centering
    \includegraphics[width=.9\linewidth]{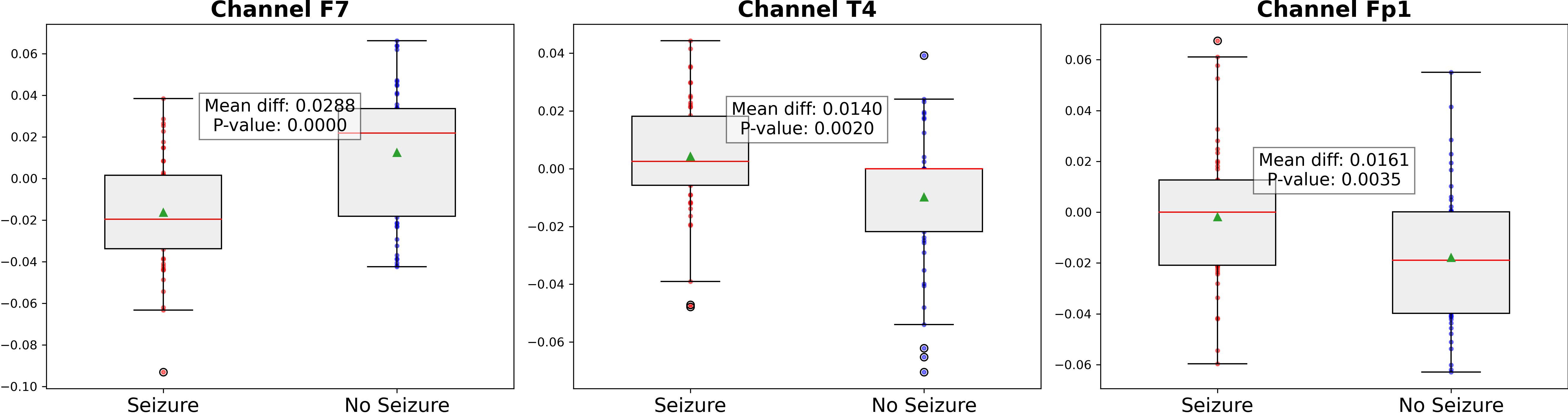}
    \caption{Channels with most significant shifts in the ranking scores between seizure free epochs and seizure epochs. For subject 5.}
    \label{fig:eeg_ranking_dynamics_subject_5}
\end{figure}

Following the approach presented in Section \ref{subsec:hierarchical_structure}, we estimated the causal influences among brain channels for each subject across different time lags during seizure and non-seizure periods. Following the procedure in Section \ref{sec:ranking_brain_regions}, we evaluate the changes in causal influence during seizure time by estimating ranking scores for all channels across 50 seizure epochs and compare them with scores from 50 seizure-free epochs. We used a t-test to determine the statistical significance of the ranking score differences. This analysis revealed significant changes in causality-based ranking scores for several channels for each examined subject, evidenced by very small p-values. For instance, in subjects 1 and 5, we highlight the EEG channels most significantly impacted by seizures. These findings, depicted in Figures \ref{fig:eeg_ranking_dynamics_subject_1} and \ref{fig:eeg_ranking_dynamics_subject_5}, illuminate how seizures alter the hierarchical architecture of brain connectivity, offering deeper insights into their impacts and pinpointing the brain regions where seizures exert the greatest influence. These results demonstrate the efficacy of our CBTR approach in uncovering the nuanced dynamics of effective connectivity influenced by seizure events.
\vspace{-10mm}

\section{Conclusion and Discussion}
\label{sec:conclusion}

In this paper, we introduced a novel framework to investigate effective brain connectivity — a domain traditionally explored through the lens of functional connectivity via persistence homology. Our proposed approach  is unique in a sense that it integrates causal inference methods with Hodge decomposition, enabling the ranking of brain regions according to their mutual causal influences. This approach offers a novel perspective on the complex dynamics with effective brain connectivity.

Our simulation study demonstrates the robustness and accuracy of our methodology in uncovering and describing various hierarchical structures within multivariate dependence networks. Additionally, the application of our method to a dataset of neonatal EEG recordings highlights its practical value. Our analysis revealed novel insights into the impact of seizures on neonatal brain connectivity, specifically identifying changes in the hierarchical positions of specific brain regions in response to seizure activity.

Our future work entails deeper investigations on epileptic seizure using our proposed method. Given its potential to enhance the localization of seizure origins, this future direction holds promise for making significant contributions to seizure. By refining our understanding of seizure-induced alterations in brain connectivity, we anticipate that our methodology will offer valuable tools for clinical diagnosis and the development of targeted interventions.

\bibliographystyle{splncs04}
\bibliography{references}

\end{document}